\documentclass[prl,twocolumn,amssymb,showpacs]{revtex4}

\usepackage{graphicx}
\usepackage{color}
\usepackage{epsfig}
\usepackage{latexsym}
\usepackage{bm}

\begin{document}

\renewcommand{\ni}{{\noindent}}
\newcommand{\dprime}{{\prime\prime}}
\newcommand{\be}{\begin{equation}}
\newcommand{\ee}{\end{equation}}
\newcommand{\bea}{\begin{eqnarray}}
\newcommand{\eea}{\end{eqnarray}}
\newcommand{\nn}{\nonumber}
\newcommand{\bk}{{\bf k}}
\newcommand{\bQ}{{\bf Q}}
\newcommand{\q}{{\bf q}}
\newcommand{\s}{{\bf s}}
\newcommand{\bN}{{\bf \nabla}}
\newcommand{\bA}{{\bf A}}
\newcommand{\bE}{{\bf E}}
\newcommand{\bj}{{\bf j}}
\newcommand{\bJ}{{\bf J}}
\newcommand{\bs}{{\bf v}_s}
\newcommand{\bn}{{\bf v}_n}
\newcommand{\bv}{{\bf v}}
\newcommand{\la}{\langle}
\newcommand{\ra}{\rangle}
\newcommand{\dg}{\dagger}
\newcommand{\br}{{\bf{r}}}
\newcommand{\brp}{{\bf{r}^\prime}}
\newcommand{\bq}{{\bf{q}}}
\newcommand{\hx}{\hat{\bf x}}
\newcommand{\hy}{\hat{\bf y}}
\newcommand{\bS}{{\bf S}}
\newcommand{\cU}{{\cal U}}
\newcommand{\cD}{{\cal D}}
\newcommand{\bR}{{\bf R}}
\newcommand{\pll}{\parallel}
\newcommand{\sumr}{\sum_{\vr}}
\newcommand{\cP}{{\cal P}}
\newcommand{\cQ}{{\cal Q}}
\newcommand{\cS}{{\cal S}}
\newcommand{\ua}{\uparrow}
\newcommand{\da}{\downarrow}
\newcommand{\red}{\textcolor {red}}

\def\lsim {\protect \raisebox{-0.75ex}[-1.5ex]{$\;\stackrel{<}{\sim}\;$}}
\def\gsim {\protect \raisebox{-0.75ex}[-1.5ex]{$\;\stackrel{>}{\sim}\;$}}
\def\lsimeq {\protect \raisebox{-0.75ex}[-1.5ex]{$\;\stackrel{<}{\simeq}\;$}}
\def\gsimeq {\protect \raisebox{-0.75ex}[-1.5ex]{$\;\stackrel{>}{\simeq}\;$}}


\title{ Nonequilibrium steady states in contact: \\ Approximate thermodynamic
structure and zero-th law for driven lattice gases }

\author{ Punyabrata Pradhan, Christian P. Amann and Udo Seifert }

\affiliation{ II. Institut f\"ur Theoretische Physik, Universit\"at Stuttgart,
Stuttgart 70550, Germany }

\begin{abstract}
\noindent{ We explore driven lattice gases for the existence of an
intensive thermodynamic variable which could determine
``equilibration'' between two nonequilibrium steady-state systems kept
in weak contact. In simulations, we find that these systems
satisfy surprisingly simple thermodynamic laws, such as the zero-th law
and the fluctuation-response relation between the particle-number
fluctuation and the corresponding susceptibility remarkably well.
However at higher densities, small but observable
deviations from these laws occur due to nontrivial contact dynamics
and the presence of long-range spatial correlations. }

\typeout{polish abstract}
\end{abstract}


\maketitle


Among the wide class of nonequilibrium systems, an important and
ubiquitous subclass are those which have a nonequilibrium steady
state (NESS). Unlike in equilibrium, a system in a NESS has a
steady current but its macroscopic properties, like in
equilibrium, are still independent of time. In contrast to
equilibrium systems, there is no well founded thermodynamic theory
even for this conceptually simplest class of nonequilibrium
systems. Intensive studies attempting to construct a suitable
statistical mechanical framework where macroscopic properties and
thermodynamic states may be characterized in a simple way have not
yet converged to a universal picture \cite{Eyink1996,
Oono_Paniconi1998, Bertini_etal, Bodineau_Derrida, Sasa2006}.

At the heart of equilibrium thermodynamics is the zero-th law
which is a consequence of equalization of intensive thermodynamic
variables when two systems are in contact. For example, when two
systems with the same temperature are allowed to exchange
particles with the total number of particles conserved, the final
equilibrium state is determined by equalization of the chemical
potentials of the two, obtained by minimizing the total free 
energy. For NESSs, we ask the same: What happens if two
NESSs are brought into contact?

Recently, there have been attempts to define an intensive
thermodynamic variable for systems such as driven granular systems
\cite{Wang_Menon}, static granular assemblies of blocked states
formed by weak driving \cite{Henkes} and a class of exactly
solvable models motivated by inelastic granular collisions
\cite{Shokef}. More generally, there has been a prescription to
define such a variable for systems in NESSs  by invoking a
hypothesis, called the asymptotic factorization property,
which has been shown to be satisfied for a class of systems having
short-range spatial correlations \cite{Bertin_Dauchot_Droz}.

For driven diffusive systems like the paradigmatic stochastic
lattice gases \cite{Zia, KLS}, which have long-range spatial
correlations, the situation is less clear. Previously, motivated
by equilibrium thermodynamics which has a rigorous basis in terms
of the large-deviation principle (LDP) \cite{Touchette}, a
hypothesis of the existence of LDP has been put forward for these
systems \cite{Eyink1996}, but not yet rigorously established.
By operationally defining a pressure and a chemical
potential, a numerical study
\cite{Hayashi_Sasa2003} indicates that a Maxwell relation is
satisfied and there may indeed exist a large-deviation function
analogous to the equilibrium free energy.

In equilibrium, the existence of an intensive variable hinges
crucially on the local thermodynamic properties of a system, i.e.,
if the system is divided into subsystems large compared to the
microscopic scales, the fluctuations in the individual subsystems
are independent of each other as a consequence of the short-range
spatial correlations in the system. In contrast, the driven
systems have generic long-range spatial correlations
\cite{Dorfman}. In this situation, it is not obvious that the
system could be divided into independent subsystems and intensive
variables analogous to those in equilibrium could be defined.

In this paper we explore by simulations the ``equilibration''
between two driven lattice gases upon contact.
Interestingly, we find that, to a very good
approximation, there is an intensive variable, like
equilibrium chemical potential, which determines the final steady
state while two such driven systems are allowed to
exchange particles. Concomitantly, the zero-th law of
thermodynamics is satisfied remarkably well. Moreover, a
fluctuation-response relation between the fluctuations in
particle-number and the corresponding susceptibility
is also well satisfied. However, at higher densities,
there are small but observable deviations from these simple
thermodynamic laws due to nontrivial contact dynamics
and the presence of long-range spatial correlations.

\begin{figure}
\begin{center}
\leavevmode
\includegraphics[width=7.0cm,angle=0]{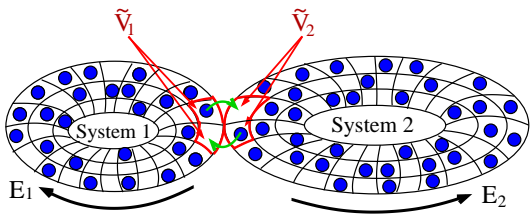}
\caption{A schematic diagram of two nonequilibrium steady states
with contact region $\tilde{V}_1$ and $\tilde{V}_2$. }
\label{dia_contact}
\end{center}
\end{figure}

We consider two systems of volume $V_1$ and $V_2$,
connected at a finite set of points
$\tilde{V}_1$ and $\tilde{V}_2$ which are subsets of $V_1$ and $V_2$
respectively, with $\tilde{V}_1, \tilde{V}_2 \ll V_1, V_2$ (see
Fig. \ref{dia_contact}). The two systems can interact and exchange
particles with each other only at the contact. The energy
$H$ of the two systems combined is given by
$\nonumber H= K_1 \sum \eta({\bf r_1}) \eta({\bf r_1}') + 
K_2 \sum \eta({\bf r_2}) \eta({\bf r_2}')  + 
\tilde{K} \sum \eta({\bf \tilde{r}_1}) \eta({\bf \tilde{r}_2})$
where sums are over nearest-neighbor sites
with ${\bf r_1}, {\bf r_1}' \in V_1$, ${\bf \tilde{r}_1} \in
\tilde{V}_1$ and ${\bf r_2}, {\bf r_2}' \in V_2$, ${\bf
\tilde{r}_2} \in \tilde{V}_2$. A site ${\bf r}$ can be occupied by
at most one particle and the occupation variable $\eta({\bf r})$
is $1$ or $0$ if the site is occupied or unoccupied, respectively.
$K_1$, $K_2$ and $\tilde{K}$ are the interaction strengths among 
particles for system 1 and 2, and at the contact, respectively. In
the simulations, we consider two-dimensional systems
($V=L \times L$) with periodic boundaries in both directions.
We choose the jump rate $w(C'|C)$ from a configuration $C$ to $C'$ 
according to the local detailed balance condition
\cite{KLS}: the jump rate from a site ${\bf r}$ to its unoccupied
nearest neighbor ${\bf r}'$ obeys $w(C'|C) = w(C|C') \exp[- \Delta
H + E (x'-x)]$ where $\Delta H = H(C') - H(C)$, $E$ is the driving
field along the $x$-direction, and $x$ and $x'$ are $x$-components
of ${\bf r}$ and ${\bf r}'$ ($k_B T =1$, $k_B$ the Boltzmann 
constant, $T$ temperature).

We choose $E=E_1$ when ${\bf r}, {\bf r}' \in V_1$,
$E=E_2$ when ${\bf r}, {\bf r}' \in V_2$ and $E=0$ otherwise.
There is {\it no} driving field along the bonds connecting
the two systems. We choose $K_1, K_2 > 0$
and $E_1, E_2$ so that systems are in the disordered (fluid)
phase \cite{Zia}. For $E_1=E_2=0$,
the combined system has the equilibrium Boltzmann distribution $
\sim \exp[- H(C)]$. For $E_1, E_2 \ne 0$, there are
currents in the steady states and the steady state
distribution is, in general, unknown. Driven bilayer systems were
studied previously \cite{Bilayr_systems} where particles
jump from one layer to the other at any site as opposed to the
case here with possibility of particle transfer only
at a small contact area.

\begin{figure}
\begin{center}
\leavevmode
\includegraphics[width=8.2cm,angle=0]{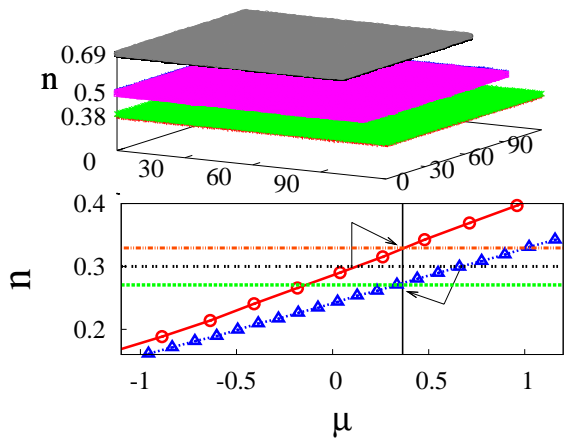}
\caption{Bottom panel: The plot of density $n$ {\it vs.} 
chemical potential $\mu$ for two $120 \times 120$ systems, 
NESS$_1$ with $K=1$, $E=2$ (circles) and NESS$_2$ with $K=E=2$ 
(triangles) with $\tilde{K}=0$. Arrows
indicate how the density changes if NESS$_1$ and NESS$_2$ with the
same initial density $n=0.30$ (denoted by middle horizontal line)
are brought into contact, reaching respective final densities $n
\simeq 0.33$ (denoted by top horizontal line) and $n \simeq 0.27$
(denoted by bottom horizontal line) with equal chemical potential
$\mu \simeq 0.36$. Top panel: Numerical experiments to test zero-th 
law ($\tilde{K}=0$) - (1) NESS$_1$ ($K=4$, $E=6$, $L=120$) with 
density $n_1$ (bottom red profile) equilibrated with NESS$_2$ 
($K=E=2$, $L=110$) with density $n_2$ (middle blue profile), 
(2) NESS$_2$ with density
$n_2$ (middle magenta profile) equilibrated with EQ$_1$
($K=1$, $E=0$, $L=100$) with density $n_3$ (top grey profile), 
and (3) NESS$_1$ with density $n_1'$ (bottom
green profile) equilibrated with EQ$_1$ with density $n_3'$ (top black
profile) where $n_1' \approx n_1$ and $n_3' \approx n_3$. } 
\label{2ndLaw_2D}
\end{center}
\end{figure}

We first report an a priori surprising observation suggesting an
effective zero-th law for systems in NESS.
When two systems are brought into contact, after relaxation involving 
exchange of particles, in the final ``equilibrated'' steady state there
is no net current across  the contact region. This property allows
an operational definition of a chemical potential of a NESS as
follows. A driven system is brought into contact with an
equilibrium system EQ whose chemical potential $\mu$ is known as a
function of density. In the final steady state, the chemical
potential $\mu$ of EQ is assigned to the driven system. We choose
a system of noninteracting hard-core particles as EQ with
density being $n_0$ and $\mu = -
(\partial s/\partial n_0) = \ln [n_0/(1-n_0)]$ with $s = -[n_0 \ln
n_0 + (1-n_0) \ln (1-n_0)]$ the equilibrium entropy per lattice
site. By varying $n_0$ of EQ in contact with a
NESS in consideration, one can get the density versus chemical
potential curve for the NESS as shown in Fig. \ref{2ndLaw_2D} 
(bottom panel). The surprising observation is that, if two NESSs,
NESS$_1$ and NESS$_2$ (chosen such that they have operationally the same
$\mu$ but different densities), are brought together, the respective 
densities do not change upon contact. Moreover, if we bring together 
two NESSs with the same density but different $\mu$, particles will
flow from the higher to lower chemical potential till the
respective densities correspond to the same
$\mu$ as indicated with arrows in Fig. \ref{2ndLaw_2D} (bottom
panel).

Thus, if two systems are separately equilibrated with a
common system with a fixed density, they will also be equilibrated
amongst themselves. Consider, e.g., two systems NESS$_1$ and NESS$_2$ kept 
in contact and having two equilibrated density profiles, 
with density $n_1$ and density $n_2$, respectively. Then, a third system 
EQ$_1$ is separately brought into contact with NESS$_2$ and the density of 
EQ$_1$ is tuned to $n_3$ such that NESS$_2$ keeps its density $n_2$ 
unchanged in the equilibrated state. Now, if NESS$_1$ with density $n_1$ 
and EQ$_1$ with density $n_3$ are brought into contact, the two density 
profiles remain almost unchanged, confirming the zero-th law 
(see Fig. \ref{2ndLaw_2D}, top panel and the explanations in the caption).

These systems are indeed far away from equilibrium since the numerical 
values of the currents in NESS$_1$ and NESS$_2$ in the bottom panel of 
Fig. \ref{2ndLaw_2D}  
are approximately $2/3$ and $1/3$ of the respective maximum currents. 
Likewise, in the top panel, NESS$_2$ with density $\approx 0.5$ has a 
homogeneous
disordered state in contrast to the corresponding equilibrium system,
with $K=2$, $E=0$ and the same density, which has a symmetry-broken phase
with different sub-lattice densities \cite{Zia}.

\begin{figure}
\begin{center}
\leavevmode
\includegraphics[width=8.2cm,angle=0]{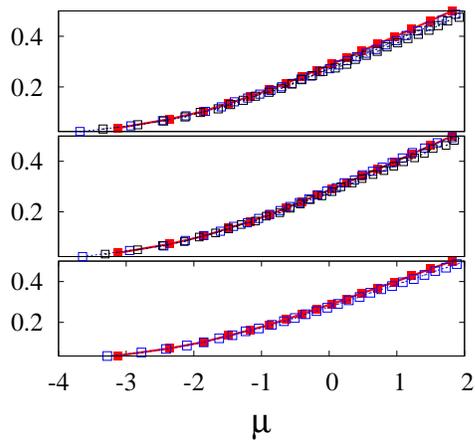}
\caption{Density $n$ {\it vs.} chemical potential $\mu$ is plotted
for a $120 \times 120$ NESS$_1$ ($K=1$, $E=2$) for the
following cases. Bottom panel: NESS$_1$ separately in contact with
$120 \times 120$ systems with EQ$_1$ ($K=E=0$, red) and EQ$_2$
($K=1$, $E=0$, blue). Middle panel: NESS$_1$ separately in contact
with EQ$_1$ (blue) and EQ$_2$ (black) with $4 \times 1$ contact area. 
Top panel: NESS$_1$ separately in contact with EQ$_1$ (blue) and 
EQ$_2$ (black) for nonzero interaction strength $\tilde{K}=1$. 
In middle and top panel, $n$ {\it vs.}
$\mu$ plot is compared with that obtained for NESS$_1$ in
contact with EQ$_1$ with $2 \times 2$ contact area. } 
\label{Zero-thLaw1}
\end{center}
\end{figure}

The chemical potential of a system may 
also be measured by keeping it in contact with any other equilibrium
system, not necessarily a noninteracting hardcore one. In Fig.
\ref{Zero-thLaw1}, we have plotted densities versus chemical
potentials for a system separately in contact with systems with
different contact area as well as nonzero interaction strength $\tilde{K}$,
the result being in good agreement with the zero-th law.

The existence of a zero-th law would be a consequence of a
putative large-deviation principle (LDP) \cite{Touchette}. To
elucidate it briefly, let us consider two systems which can
exchange particles such that $N_1+N_2 = N = const$, $N_1$,
$N_2$ the number of particles in systems 1 and 2 respectively.
Assuming that the LDP holds, the probability $P(N_1, N_2)$ of a large
deviation in $N_1, N_2$ is given by $ P(N_1, N_2) \sim \left[
e^{V_1 s_1(n_1)} e^{V_2 s_2(n_2)} \right] e^{-S(N)} $ in the limit
of $N_1, N_2, V_1, V_2 \gg 1$ where $n_1=N_1/V_1$
and $n_2=N_2/V_2$ being finite and $\exp[-S(N)]$ the
normalization constant (`$\sim$' implying equality in terms of logarithm). 
The functions $s_1(n_1)$, $s_2(n_2)$ are
called the large-deviation functions (LDF).
In writing so, the correlation between
systems has been neglected as a boundary-effect in the
limit of large volume. This assumption of a product measure of
$P(N_1, N_2)$ essentially implies that the LDFs $s_1(n_1)$ and
$s_2(n_2)$ are local function of the respective densities. The
macroscopic state, under the constraint $N_1 + N_2 = const$,
is determined by maximizing $\ln P(N_1, N_2)$ where
the chemical potentials $\mu_1= - \partial s_1/\partial n_1$ and
$\mu_2= - \partial s_2/\partial n_2$ being equal in the final
steady state. Clearly the consequence of the LDP 
is a zero-th law as presented in Fig. \ref{2ndLaw_2D}. 
Another interesting consequence of a putative LDP would be a
relation between the susceptibility and the fluctuation in
particle-number of the system 1 in a NESS when it is in contact with
the system 2 being a large reservoir characterized by a chemical 
potential $\mu$.
Then, one gets the following fluctuation-response relation as
in equilibrium, \be \chi \equiv \frac{\partial \langle N_1
\rangle}{\partial \mu} = (\langle N_1^2 \rangle - \langle N_1
\rangle^2) \equiv \sigma_{N_1}^2 \label{FR1}. \ee We first 
proceed to test this relation for a NESS in contact with an
equilibrium reservoir with density $n_0$, consisting of
noninteracting hardcore particles with $\mu = - (\partial
s/\partial n_0) = \ln [n_0/(1-n_0)]$. For better numerical
accuracy, we check the integrated version of Eq. \ref{FR1} by
defining the integrated susceptibility $I_{\chi}(\mu) \equiv
\int^{\mu}_{\mu_0} (\partial \langle N_1 \rangle/\partial \mu) d\mu = \langle
N_1(\mu) \rangle - \langle N_1(\mu_0) \rangle$ and the integrated
fluctuation $I_{\sigma}(\mu) \equiv \int^{\mu}_{\mu_0}
(\sigma_{N_1}^2) d \mu$.

\begin{figure}
\begin{center}
\leavevmode
\includegraphics[width=8.0cm,angle=0]{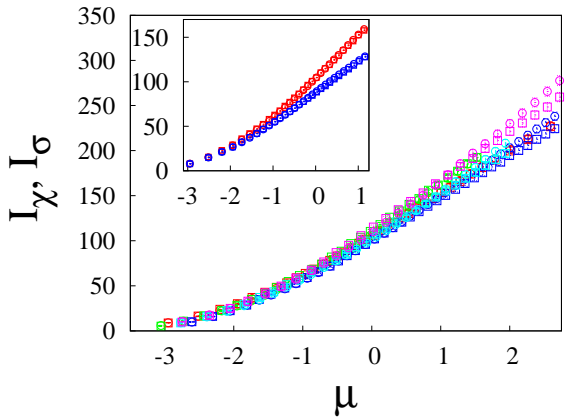}
\caption{Integrated susceptibilities $I_{\chi}$ (squares) and
fluctuations $I_{\sigma}$ (circles) ${\it vs.}$ 
chemical potential $\mu$ are plotted for a $20 \times 20$
NESS$_1$ with $K=1$, $E=2$, separately in contact with
five different $100 \times 100$ reservoirs: (1) $K=E=0$ (red),
(2) $K=1$, $E=2$ (green), (3) $K=1$, $E=4$ (blue), 
(4) $K=2$, $E=4$ (sky-blue), (5) $K=2$, $E=6$ (magenta). 
Inset: Same quantities are plotted
for two different $20 \times 20$ systems, NESS$_1$
with $K=1$, $E=2$ (blue) and NESS$_2$ $K=E=2$ (red), in
contact with a reservoir ($K=E=0$, $L=100$). }
\label{eq_noneq_reservoirs2}
\end{center}
\end{figure}

We take a $20 \times 20$ nonequilibrium system NESS$_1$ with
$K=1$, $E=2$ and keep it in contact with a $100 \times 100$
equilibrium reservoir of noninteracting hardcore particles
($K=E=0$), called RES$_1$. Then we vary the chemical potential
$\mu$ (or equivalently the density $n_0$) of RES$_1$ in
small steps from an initial value $\mu_0=-3.5$ and calculate
$\sigma_{N_1}^2$ for each value of $\mu$. We repeat this procedure
by keeping NESS$_1$ separately in contact with various other
reservoirs of size $100 \times 100$ whose chemical potentials can
be measured by keeping these reservoirs in contact with the
RES$_1$. In Fig. \ref{eq_noneq_reservoirs2}, we plot
$I_{\chi}(\mu)$ and $I_{\sigma}(\mu)$ as a function of $\mu$.
Provided that the Eq. \ref{FR1} is valid, all the curves should
fall on each other. Up to chemical potential $\mu \simeq 1$, we
observe a quite good collapse within the numerical accuracy. We
also consider two different systems, NESS$_1$ with $K=1$,
$E=2$ and NESS$_2$ with $K=E=2$, separately
in contact with RES$_1$. In the inset of Fig.
\ref{eq_noneq_reservoirs2}, we plot $I_{\chi}(\mu)$ and
$I_{\sigma}(\mu)$ which are in good agreement with the
fluctuation relation in Eq. \ref{FR1}.

\begin{figure}
\begin{center}
\leavevmode
\includegraphics[width=8.0cm,angle=0]{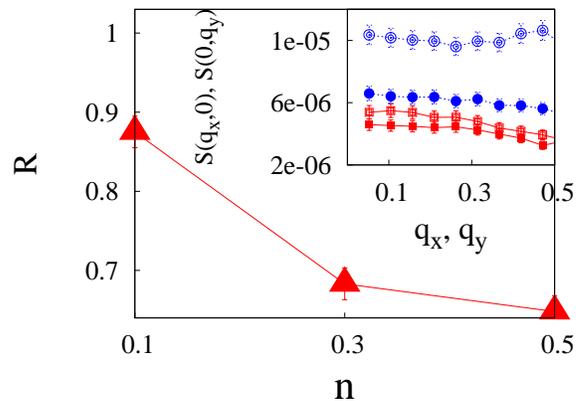}
\caption{The ratio $R=[\lim_{q_y \rightarrow 0}
S(0,q_y)]/[\lim_{q_x \rightarrow 0} S(q_x, 0)]$ {\it vs.} density $n$
is plotted for a NESS with $K=1$, $E=2$, $L=120$. 
Inset: The structure factors $S(q_x,0)$ (open
points) and $S(0,q_y)$ (filled points) {\it vs.} $q_x$ and $q_y$
respectively are plotted for densities $n=0.1$ (red) and
$n=0.5$ (blue). } \label{struc_fact}
\end{center}
\end{figure}

At higher chemical potentials, there are observable deviations
from this simple thermodynamic behavior. In Fig. 
\ref{eq_noneq_reservoirs2}, the $I_{\chi}$ {\it vs.} $\mu$ and 
$I_{\sigma}$ {\it vs.} $\mu$ curves do not fall on each
other for $\mu \gsim 1$. Correspondingly, in this density regime,
the zero-th law does not hold strictly as seen in Fig.
\ref{Zero-thLaw1}. However, these violations are not simply due to
a finite-size effect and persist for much larger system sizes. To
investigate the possible reasons for the violations, we also study the
behavior of spatial density correlation functions for various
densities. Unlike equilibrium systems, the nonequilibrium systems,
due to the presence of a driving field, are expected to have
generic long-ranged spatial correlations 
\cite{Zia, Dorfman, Garrido, Grinstein}, arising because the structure 
factor $S(q_x, q_y)$, i.e., the Fourier transform
of the spatial density correlation function, becomes singular when
$q_x, q_y \rightarrow 0$ with $R=[\lim_{q_y \rightarrow 0}
S(0,q_y)]/[\lim_{q_x \rightarrow 0} S(q_x, 0)] \ne 1$. This gives
rise to the long-range spatial correlations decaying as $A/r^d$
with distance $r$ in $d$ dimension with the amplitude $A \propto (R-1)$ 
\cite{Zia}. In Fig.
\ref{struc_fact}, we have plotted the ratio $R$ versus density $n$
and the structure factors $S(q_x, 0)$ and $S(0, q_y)$ for two
different densities in inset of Fig. \ref{struc_fact} for a NESS 
with $K=1$, $E=2$, $L=120$. The ratio $R$ deviates from $1$ more strongly
with increasing density, indicating an increase of long-range
correlations which manifest themselves through the nonlocal effect 
of the contact. In Figs. \ref{Zero-thLaw1} and \ref{eq_noneq_reservoirs2}, 
collapse of various $n$ {\it vs.} $\mu$ and $I_{\sigma}$ {\it vs.} $\mu$ 
(or $I_{\chi}$ {\it vs.} $\mu$) curves for different contact dynamics 
are not very good for $\mu \gsim 1$. Clearly, 
the effect of the contact is felt throughout the systems and
this nontrivially changes the corresponding thermodynamic
properties. Therefore, the break-down of an exact equilibrium-like
structure or, in other words the break-down of the
product-measure assumption in the LDP, indicates the important role
of the contact dynamics \cite{Bertin_Dauchot_Droz} and the long-range 
correlations in a driven system.

In summary, our numerical study of coupled driven lattice gases
has revealed a surprisingly simple thermodynamic structure with an
effective zero-th law like behavior concerning exchange of
particles and the corresponding fluctuation-response relation.
This thermodynamic structure is not exact since
there are small but observable deviations at higher densities. Their
physical origin is rooted in the nontrivial contact dynamics and
the presence of long-range spatial correlations
which invalidate the asymptotic factorization property.
As an open question, it would be interesting to see 
whether systems with more than one conserved quantity
(e.g., models with two species) exhibit a similar behavior. Finally
our study prompts the question whether such an approximate
thermodynamic structure is typical just for driven lattice gases
or generically occurs in other coupled NESSs with long-range
correlations as well.

We thank J. Krug and R. K. P. Zia for discussions.



\end{document}